\documentclass{aa}
\usepackage[dvips]{graphics}
\usepackage{latexsym}

\def\gtsima{$\; \buildrel > \over \sim \;$}
\def\ltsima{$\; \buildrel < \over \sim \;$}
\def\prosima{$\; \buildrel \propto \over \sim \;$}
\def\gsim{\lower.5ex\hbox{\gtsima}}
\def\lsim{\lower.5ex\hbox{\ltsima}}
\def\simgt{\lower.5ex\hbox{\gtsima}}
\def\simlt{\lower.5ex\hbox{\ltsima}}
\def\simpr{\lower.5ex\hbox{\prosima}}

\begin{document}

\thesaurus{03(11.05.1;11.05.2;11.06.1;11.19.3;12.12.1;11.03.1)}

\title{Detection of Strong Clustering of Extremely Red Objects:
Implications for the Density of \boldmath{$z>1$} Ellipticals}

\author{
	E. Daddi \inst{1} 
	\and
	A. Cimatti \inst{2}
	\and
	L. Pozzetti \inst{2,3}
	\and
	H. Hoekstra \inst{4}
	\and
	H.J.A. R\"ottgering \inst{5}
	\and
	A. Renzini \inst{6}
	\and 
	G. Zamorani \inst{3}
	\and
	F. Mannucci \inst{7}
}
\institute{ 
	Universit\`a degli Studi di Firenze, 
	Dipartimento di Astronomia e Scienza dello Spazio,
	Largo E. Fermi 5, I-50125 Firenze, Italy
 	\and  
	Osservatorio Astrofisico di Arcetri,
	Largo E. Fermi 5, I-50125 Firenze, Italy
	\and
	Osservatorio Astronomico di Bologna,
	Via Ranzani 1, I-40127 Bologna, Italy
	\and
	Kapteyn Institute, Postbus 800, 9700 AV Groningen, The Netherlands
	\and
	Sterrewacht Leiden, Postbus 9513, 2300 RA Leiden, The Netherlands
	\and
	European Southern Observatory, D-85748 Garching, Germany
	\and
	CAISMI--CNR, Largo E. Fermi 5, I-50125 Firenze, Italy
} 

\offprints{edaddi@arcetri.astro.it\\ Partially based on observations made at the European
Southern Observatory in La Silla, Chile.}
\date{Received ; accepted }

\maketitle
\markboth{Daddi et al.: Clustering of EROs}{Daddi et al.: Clustering of EROs}

\begin{abstract}

We present the results of a wide--field survey for extremely red objects
(EROs hereafter), the widest so far, based on $Ks$ and $R$ band imaging. 
The survey covers 701 arcmin$^2$  and it is 85\% complete to $Ks\leq18.8$
over the whole area and to $Ks\leq19.2$ over 447.5 arcmin$^2$.
Thanks to the wide field covered, a complete sample 
of about 400 EROs with $R-Ks\geq$5 was selected. 
The distribution of the EROs on the sky is strongly 
inhomogeneous, being characterized by overdensities and large voids. 
We detect at the $8\sigma$ level a strong clustering signal of the EROs
which is about an order of 
magnitude larger than the clustering of $K$-selected field galaxies in 
the same magnitude range. 
A smooth trend of increasing clustering amplitude with the $R-Ks$ color is
observed.
These results are strong evidence that the largest fraction of EROs 
is composed of high--$z$ ellipticals, of which we detect for the first time the
$z\simgt1$ large scale structure clustering signal.
We show how the surface density variations of 
the ERO population found in our survey can explain the highly
discrepant results obtained so far on the density of $z>1$ ellipticals, 
and we briefly discuss the main implications of our results for the 
evolution of elliptical galaxies. The number counts and the colors of the 
$K$-selected field galaxies are also presented and briefly discussed.


\end{abstract}

\section{Introduction}

Near-infrared surveys prompted the discovery of a population of objects 
with very red optical-infrared colors (Extremely Red Objects, EROs hereafter; 
e.g. Elston et al. 1988, McCarthy et al. 1992, Hu \& Ridgway 1994; Thompson
et al. 1999; Yan et al. 2000; Scodeggio \& Silva 2000). In general,
objects have been classified as EROs when they had redder colors than 
late type galaxies (with negligible dust extinction) at any redshift.
However, depending on the depth of the photometry and on the available 
filters, different selection criteria have been used to select EROs. In 
this paper, EROs are defined as objects with $R-Ks\geq$5 (see Sect. 5 for 
more details on this choice). 

The red colors of EROs are consistent with two classes of galaxies: they 
could be old, passively evolving elliptical galaxies at $z\simgt1$ which are so 
red because of the large $K$--correction. 
EROs may also be strongly dust--reddened star--forming galaxies 
or AGN. The observational results of the last few years showed that both 
classes of galaxies are indeed present in the ERO population: on one hand,
a few objects were spectroscopically confirmed to be  $z>1$ ellipticals 
(Dunlop et al. 1996, Spinrad et al. 1997, Liu et al. 2000, and
marginally, Soifer et al. 1999), or to have surface brightness profiles
consistent with being dynamically relaxed early type galaxies (e.g. 
Stiavelli et al. 1999, Benitez et al. 1999). On the other hand, other
EROs have been detected in the sub--mm (Cimatti et al. 1998, Dey et 
al. 1999, Smail et al. 1999, Andreani et. al. 2000), thus providing examples of high--redshift 
starburst galaxies reddened by strong dust extinction and characterized
by high star formation rates. The relative contribution of the two classes 
of objects to the whole ERO population is still unknown, but there are 
preliminary indications, based on near-IR and optical spectroscopy and on 
surface brightness analysis, that ellipticals may represent the largest 
fraction of this population
(e.g. Cimatti et al. 1999, 2000; Liu et al. 2000; Moriondo et 
al. 2000). 
A small fraction of low-mass-stars and brown dwarfs among EROs is also expected 
in case of unresolved objects (e.g. Thompson et al. 2000, Cuby et al. 1999).

The importance of studying EROs is clear especially for the clues that
they could provide on the formation and evolution of elliptical galaxies. 
For instance, existing realizations of hierarchical models of galaxy 
formation predict a significant
decline in the comoving density of the ellipticals with $z$, as they should form
through merging at $z\simlt2$ (Kauffmann 1996, Baugh et al. 1996a), 
so that a measure of a decline of their comoving density would provide a stringent 
proof of these models. Conflicting results have been found so far about
such issue: some works claim the detection of a deficit of $z>1$ 
ellipticals (e.g. Kauffmann et al. 1996, Zepf 1997, Franceschini 
et al. 1998, Barger et al. 1999, Menanteau et al. 1999), whereas others find 
that a constant comoving density of ellipticals up to $z\sim2$ is consistent
with the data (Totani \&
Yoshii 1997, Benitez et al. 1999, Broadhurst \& Bowens 2000, Schade et al. 
1999). 
A potentially serious problem in these studies is suspected to be the
influence of the field-to-field variations in the density of EROs 
due to the small fields of view usually covered in the near-infrared.
For instance, Barger et al. (1999) found a very low surface density of 
EROs in a 60 arcmin$^2$ survey to $K=20$, while on a similar area 
McCracken et al. (1999) observed a density three times larger at the
same $K$ level.

Similar uncertainties have been found in the attempts of deriving the fraction
of high redshift galaxies among IR selected samples, which is believed to
be a stringent test for the formation of the massive galaxies (Broadhurst
et al. 1992, Kauffmann \& Charlot 1998). Fontana et al. (1999), 
using photometric redshifts, found that the fraction of high--$z$ galaxies
in a collection of small deep fields complete to $K=21$ was low and comparable
with the predictions of the cold dark matter (CDM) models, but not with passive evolution (PLE) models. 
The preliminary results of Eisenhardt et al. (2000), suggesting a much 
higher fraction of high--$z$ galaxies, are instead
consistent with both CDM and PLE models.

The main aim of our survey was to encompass the
difficulties induced by the cosmic variance,
obtaining a sample of EROs on a large area, at 
moderately deep $K$ levels, in order to minimize and possibly to detect 
the effects of their clustering, and to compare the observed density with 
that expected in the case of passive evolution of ellipticals. 
Our survey is larger by more than a factor of four than the Thompson
et al. (1999) survey, and by more than an order of magnitude than all the other
previous surveys for EROs, at the same limiting magnitudes.
With the large area covered we aimed also to detect a sample,
or place limits to the surface density of the very rare class of extreme EROs
with $R-Ks\geq7$.

In this paper, the observational results of this survey and their main
implications are presented. A more detailed interpretation of our findings 
will be presented in a forthcoming paper (Daddi et al., in preparation).

The paper is organized as follows: we first describe the data reduction 
and analysis, then we present the counts of field galaxies. 
In Sect. \ref{sec:eros} the sample of EROs is described. Sect. \ref{sec:clustering} 
contains the analysis of the clustering of field galaxies and EROs. The 
main implications of our findings are discussed in Sect. \ref{sec:implic}.
H$_{\rm 0}=50$~km~s$^{-1}$~Mpc$^{-1}$ throughout the paper.

\section{Observations, data reduction and photometry}

\subsection{$Ks$-band imaging}
\label{sec:Ks}

The $Ks$ observations were made with the ESO NTT 3.5m telescope in La Silla, 
during the nights of 27--30 March 1999, using the SOFI camera (Moorwood et al. 
1998) with a field of view of about 5$\arcmin \times $5$\arcmin$.
SOFI is equipped with a Hawaii HgCdTe 1024x1024 array, with a scale of 
0.29$\arcsec$/pixel. The $Ks$ filter has $\lambda_{\rm c} = 2.16~\mu m$ and 
$\Delta\lambda \sim 0.3~\mu m$ and it is slightly bluer than the standard 
$K$ filter in order to reduce the thermal background.

The center of the observed field is at $\alpha$ = 14$^h$49$^m$29$^s$ and 
$\delta$ = 09$\degr 00\arcmin 00\arcsec$ (J2000). The observed field is
one of the fields described in Yee et al. (2000) to which we refer for
details about its selection. The main criteria were not to
have any apparent nearby clusters and to be at high galactic latitude.

The images were 
taken with a pattern of fixed offsets of 144$\arcsec$ (about half of the 
SOFI field of view) over a grid of 9$\times$13 pointings. 
The total area covered by the observations was about 24$\times$34 arcmin, 
with a local integration time of 12 minutes in the central deepest region 
of the field. In the shallower region, the effective integration time 
is reduced to not less than 6 minutes. The total amount of time
required to cover the whole field was about 5.5 hours.

The data reduction was carried out using the IRAF software. The images were 
flat-fielded with twilight flats. The sky background  was estimated
and subtracted for each frame using a clipped average of 6--8 adjacent 
frames (excluding the central frame itself). The photometric calibration 
was achieved each night with the observation of 5-7 standard stars taken 
from Persson et al. (1998). The zero-points have a scatter of $\sim$ 
0.015 magnitudes in each night and a night-to-night variation within 0.02 
magnitudes. 
Each frame was scaled to the same photometric level correcting for the 
different zero-points and airmasses. Accurate spatial offsets were 
measured for each frame using the area in common with the adjacent frames. 
The images were then combined, masking the known bad pixels, in order to 
obtain the final mosaic. The cosmic rays were detected and replaced by 
the local median using the task {\it cosmicrays} of the IRAF 
package {\it ccdred}. The effective seeing of the final coadded mosaic
ranges from 0.9$\arcsec$ to 1.1$\arcsec$.

\subsection{$R$-band imaging}

The $R$-band data were taken in May 19--21 1998 with the 4.2m William 
Herschel Telescope on La Palma. The observations were done using the 
prime focus camera, equipped with a thinned 2048x4096 pixels EEV10 chip, 
with a scale of 0.237$\arcsec$/pixel. This gives a field of view of about 
8.1$\arcmin \times 16.2\arcmin$. A standard Johnson $R$--filter was used. The whole field 
has been covered by a mosaic of 6 pointings. Each pointing consisted of at least
3 exposures of 1200s taken with small offsets. The total integration
times per pointing was therefore 3600s, with the exceptions of two 
pointings with 4800s and 6000s.

The photometric calibration
was achieved with standard stars taken from Landolt (1992) with a scatter
in the zeropoints, from
the different stars used, below 0.01 magnitudes. 
The images were de-biased and then flatfielded, using a master flatfield
constructed from the science exposures, scaled to the same zeropoint and 
then combined.
The seeing of the final $R$-band mosaic was between 0.7$\arcsec$ and 
0.8$\arcsec$.

\subsection{Sample selection and $Ks$ photometry}
\label{compl}

The software SExtractor (Bertin \& Arnouts 1996) was run on the $Ks$ 
mosaic with a background weighted threshold in order to take into account the 
depth variations across the area, as defined by SExtractor. 
Among the detected objects, all those with S/N$>5$ 
in a 2$^{\prime\prime}$ circular aperture (twice the average seeing FWHM) 
were selected and added to the catalog. A few spurious detections (e.g.
close to image defects) have been excluded after a visual inspection
of the image. The final catalog includes 4585 objects. In the central 
deepest region, the 5$\sigma$ limiting aperture magnitude is 
$Ks$(2$^{\prime\prime}$)=19.6, whereas in the remaining area such
limit is $Ks$(2$^{\prime\prime}$)\simgt19.2 because of the reduced integration 
time.

Isophotal magnitudes were measured with a limiting threshold of about
0.7$\sigma_{\rm sky}$ corresponding in the central area to  a surface 
brightness limit of about $\mu_{\rm lim}\sim21$ mag arcsec$^{-2}$. The aperture 
correction from 2$^{\prime\prime}$ to total magnitudes was estimated 
throughout the area by measuring the difference between the isophotal 
and the 2$\arcsec$ aperture magnitudes for the stars with $Ks<$16. 
A differential correction, in the range of 0.16--0.30 magnitudes, 
was measured for different regions of the mosaic with a typical scatter 
less than 0.03 magnitudes. For the bright objects the isophotal 
magnitudes were on average consistent with the Kron automatic aperture
magnitudes. However, we adopted the isophotal magnitudes 
because the Kron magnitudes are rather unstable at faint flux levels,
where the low signal often does not allow to define the correct automatic 
aperture.

The total $Ks$ magnitudes were then defined as the brightest between the 
isophotal and the corrected aperture magnitude. This allowed to safely assign 
a total magnitude for both the faint and the bright objects. 
The typical $Ks$ magnitude where the corrected aperture magnitude begins to be
adopted as the total one is $Ks$$\sim$18 in the central deepest region. 

The completeness of our catalog has been estimated by adding artificial 
objects to the $Ks$ mosaic in empty positions, using the IRAF package 
{\it artdata}. Point-like sources as well as objects with De Vaucouleurs 
and exponential profiles (convolved with the seeing PSF) were simulated,
and SExtractor was run with the same detection parameters as for the real
data.
The 85\% completeness magnitude for the deepest area is $Ks=19.2$ for 
point-like sources. The completeness decreases to $\sim$\ 70\% for the
worst case that we have tested, i.e. for 
exponential galaxies with 0.7$\arcsec$ half--light radius. 
In the shallower area the corresponding limiting magnitude is $Ks\leq18.8$.
Most of the $Ks\simgt18.5$ galaxies are anyway expected to be only barely
resolved with the $Ks$ 1$\arcsec$ seeing (Saracco et al. 1997),
and this certainly occurs for the distant $z\simgt1$ ellipticals,
and thus their completeness limits can be assumed to be similar to those 
for stars. 

\subsection{$R$-band photometry and colors}

In order to recover the $R$-band counterparts of the $Ks$-selected 
objects, a coordinate mapping between the $Ks$ and the $R$ images
was derived. 
SExtractor was then run in ASSOC mode with a search box of 2$\times$
FWHM$_{\rm R}$. The regions around bright stars or defects in the $R$ and 
$Ks$ band images were excluded from this analysis. The final effective 
area is 701 arcmin$^2$ at $Ks\leq18.8$ and 447.5 arcmin$^2$ at 
$18.8<Ks\leq19.2$.

Whenever an object had S/N$<$3 in the $R$-band image 3$\sigma$ limits were 
assigned. The 3$\sigma$ limiting magnitude in 
a 2$\arcsec$ diameter aperture is $R>26.2$ for most of the area, 
reaching $R>26.5$ in the deepest pointing. When S/N$>$3, 2$\arcsec$ 
diameter corrected aperture magnitudes were assigned to each object. 
The aperture correction was derived in the same way as for the $Ks$-band, 
with slightly smaller corrections because of the better $R$-band seeing. 
The magnitudes were dereddened for Galactic extinction. At the Galactic 
coordinates of the center of our field ($l\sim5\fdg5$,  $b\sim57\degr$), the 
extinction coefficients from Burstein \& Heiles (1982) and from Schlegel 
et al. (1998) are $A_{\rm B}=0.04$ and $A_{\rm B}=0.13$ respectively.
Since the two values are derived in different ways, neither
of the two can be discarded. The average was therefore adopted, obtaining a correction 
of 0.052 magnitudes in $R$ and negligible in $Ks$. This introduces 
an uncertainty of $\sim$ 0.03 magnitudes in the dereddened $R$ magnitudes.

\begin{figure}[ht]
\resizebox{\hsize}{!}{\includegraphics{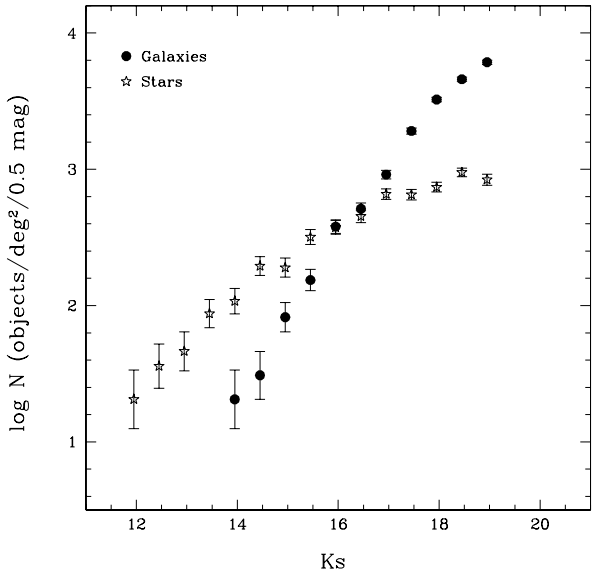}}
\caption{\footnotesize 
The observed number counts for both stars and galaxies in our survey. 
The error bars indicate the poissonian uncertainties. 
}
\label{fig:stars1}
\end{figure}

Finally, the $R-Ks$ colors are defined for all the objects as the 
difference between the $R$ and $Ks$ corrected aperture magnitudes.
Thanks to the depth of the $R$-band data,
colors as red as $R-Ks=7$ could be measured down to the $Ks$ magnitude
limits of our survey. Because of the long integrations used  
in the $R$-band the objects with $R\simlt 20$ were saturated.
{This effect is obviously more important for stars than for galaxies;}
moreover, since we are interested in the extremely red galaxy population, 
the effects due to saturation have no impact on our results.

\subsection{Star--Galaxy classification}
\label{S/G}

The star--galaxy (S/G) separation was done by means of the SExtractor 
CLASS\_STAR parameter in both the $R$ and $Ks$ band. This classification was
found to be reliable for objects with $Ks\simlt$17.5 and $R\simlt$24. 
Because of the seeing variations through the area in both bands,  
a variable CLASS\_STAR threshold was adopted in different subareas. Given the better 
seeing of the $R$-band, the classification was based mostly on that band, 
switching to the $Ks$ CLASS\_STAR for objects close to the saturation 
level in $R$. 
{Objects which are fainter than $Ks$ = 17.5 and $R$ = 24
have not been classified and have been considered to be galaxies.}
This means that a S/G separation can be
provided
only for objects with colors $R-Ks\simlt$5 at $Ks$=19, $R-Ks\simlt$6 at 
$Ks$=18, and so on (see Fig. \ref{fig:zoom} and the upper panel of 
Fig. \ref{fig:ccdiag}). {From the small number of objects classified
as stars which are close to the diagonal straight line which indicates
our S/G classification limit in the color - magnitude plane, we can safely 
conclude that our inability to properly classify very red, faint objects
has almost no effect on the total star number counts and, therefore, 
on the galaxy number counts.}
In addition, it is possible that near the faint limit
of our survey a very small fraction of
very compact, blue galaxies, such as for instance AGN or
compact narrow emission line galaxies (e.g. Koo \& Kron 1988), could have been incorrectly classified 
as stars.

\section{$Ks$-band number counts}
\label{sec:counts}

Galaxy number counts in the $Ks$ band can provide more advantages in
studying galaxy evolution and cosmological geometry than optical counts
because they are much less sensitive to the evolution of stellar
population and to the dust extinction.
Our survey, which covers the 
magnitude range $14<Ks<19.2$, represents the widest among the previous 
deep surveys at levels fainter than $Ks>18$. 

\begin{table}[ht]
\begin{flushleft}
\caption{Differential Number Counts}
\protect\label{tab:XX}
\begin{tabular}{cccc}
\noalign{\smallskip}
\hline
\noalign{\smallskip}
\multicolumn{1}{c}{$Ks$ range} & \multicolumn{1}{c}{Area$^{[2]}$} & \multicolumn{1}{c}{Galaxies} & \multicolumn{1}{c}{Stars} \\
\noalign{\smallskip}
\hline
\noalign{\smallskip}
 11.7 -- 12.2   & 701 &     -     &     4         \\ 
 12.2 -- 12.7   & " &     -     &     7          \\ 
 12.7 -- 13.2   & " &     -     &     9          \\ 
 13.2 -- 13.7   & " &     -     &    17         \\ 
 13.7 -- 14.2   & " &     4     &    21         \\ 
 14.2 -- 14.7   & " &     6     &    38         \\ 
 14.7 -- 15.2   & " &     16    &    37         \\ 
 15.2 -- 15.7   & " &     30    &    62         \\ 
 15.7 -- 16.2   & " &     74    &    73         \\ 
 16.2 -- 16.7   & " &     100   &    88         \\ 
 16.7 -- 17.2   & " &     178   &   128         \\ 
 17.2 -- 17.7   &  " &    372   &   127         \\ 
 17.7 -- 18.2   & " &     633   &   144         \\ 
 18.2 -- 18.7   & " &     892   &   185        \\ 
 & & & \\
 18.7 -- 18.8   &  " &    200    &   32    \\ 
 18.8 -- 19.2$^{[1]}$   & 447.5 &  628   &   84   \\ 
\noalign{\smallskip}
\hline
\end{tabular}
\end{flushleft}
\footnotesize{$^{[1]}$ The last bin includes only the objects in the deepest
region.\\ $^{[2]}$ arcmin$^2$}
\end{table}

Table \ref{tab:XX} summarizes the number of galaxies and stars detected in 
each $Ks$ bin and Fig. \ref{fig:stars1} shows the corresponding differential 
counts in 0.5 magnitude bins. No  correction for
incompleteness was applied.  Galaxies start to dominate over 
stars at $Ks \sim 16.5$ and their surface density is about a factor of 8
higher than the stellar surface density at $Ks \sim 19.$ 

\begin{figure}
\resizebox{\hsize}{!}{\includegraphics{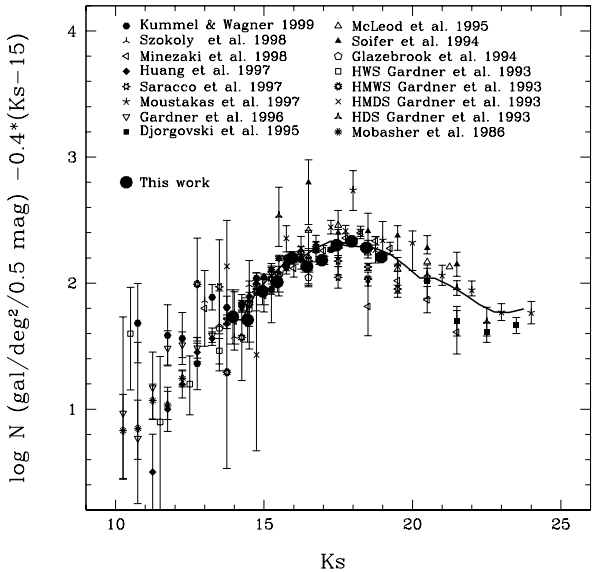}}
\caption{\footnotesize
Counts of galaxies from our survey compared to a collection of 
published data. The solid curve shows the average counts as estimated by
Hall \& Green (1998).
}
\label{fig:counts_gala}
\end{figure}

The slopes of the galaxy number counts were derived over the magnitude range 
covered by our survey. At bright magnitudes a slope of $\gamma= 0.53\pm0.02$
is found in the range $14<Ks<17.5$. 
We confirm that the $K$-band galaxy counts show a 
flattening at $Ks \sim 17.5$, where the best fit slope changes from 
$\gamma=0.53$ to $\gamma=0.32\pm0.02$ (see Fig. \ref{fig:counts_gala}). 
The leveling off of the counts below a slope of 0.4 
indicates that the differential contribution to the 
extragalactic background light (EBL) in the $K$ band
peaks at $Ks\sim17-18$ and then starts to decrease at fainter fluxes.
The contribution to the EBL over the 
magnitude range $14\leq Ks\leq19.2$ sampled by our survey is about 
$4.20$ nW/m$^2$/sr, which constitutes about $53 \%$ of the estimated
EBL from discrete 
sources in the $K$-band (cf. Pozzetti et al.  1998, Madau \& Pozzetti 2000).

Fig. \ref{fig:counts_gala} shows the differential galaxy number counts in 
our survey compared with a compilation of $K$-band published surveys.
No attempt was made to correct 
for different filters. As shown in the figure, our counts are 
in very good agreement with the average
counts of previous surveys (Hall \& Green 1998).  

\begin{table}[hb]
\begin{flushleft}
\caption{Galaxy Median Colors}
\protect\label{tab:median}
\begin{tabular}{ccc}
\noalign{\smallskip}
\hline
\noalign{\smallskip}
\multicolumn{1}{c}{$Ks$ range} & \multicolumn{1}{c}{Galaxies} & \multicolumn{1}{c}{Median $R-Ks$} \\
\noalign{\smallskip}
\hline
\noalign{\smallskip}
16.5 -- 17.0 & 143 & 3.54 \\
17.0 -- 17.5 & 277 & 3.80 \\
17.5 -- 18.0 & 522 & 3.92 \\
18.0 -- 18.5 & 759 & 4.08 \\
18.5 -- 18.8 & 613 & 4.11 \\
18.8 -- 19.2 & 628 & 4.04 \\
\noalign{\smallskip}
\hline
\end{tabular}
\end{flushleft}
\end{table}

\begin{figure}[ht]
\resizebox{\hsize}{!}{\includegraphics{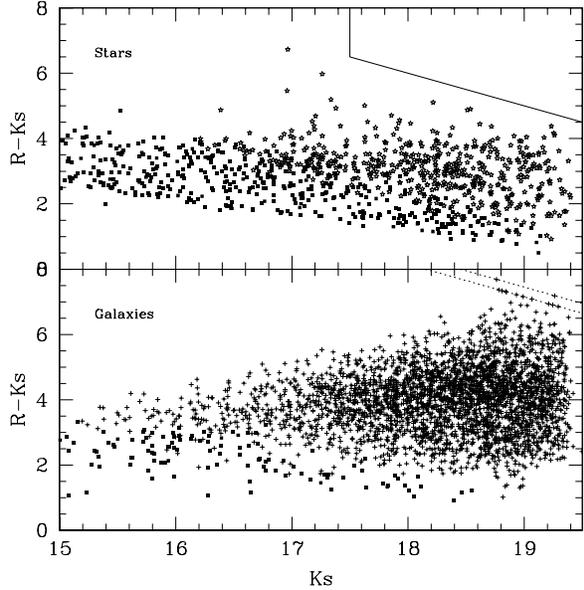}}
\caption{\footnotesize Color magnitude diagram for the stars (upper panel) 
and galaxies (lower panel) samples.
The diagonal straight line in the upper panel indicates
our S/G classification limit in the color - magnitude plane (see text). 
The bluer objects plotted with filled squares have at least one pixel
close to saturation in the $R$-band. Some of the bluest saturated galaxies at $Ks>17$ may
actually be stars. The reddest galaxies with $R-Ks\simgt7$ have
color lower limits as they are not detected in the $R$-band (see Fig. \ref{fig:zoom}).
}
\label{fig:ccdiag}
\end{figure}
\begin{figure}[ht]
\resizebox{\hsize}{!}{\includegraphics{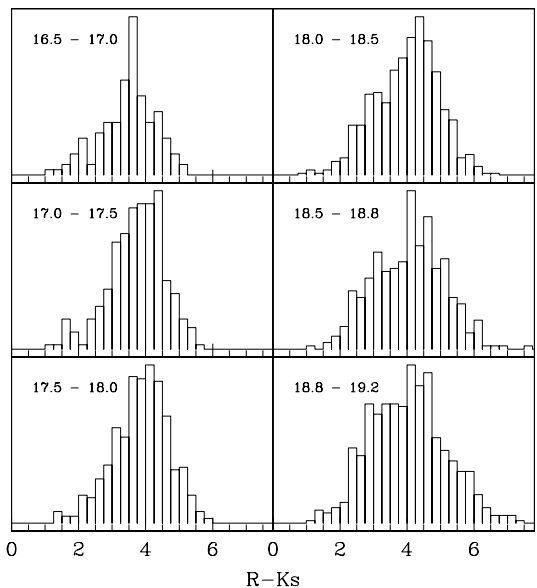}}
\caption{\footnotesize Distributions of the $R-Ks$ colors for the galaxies of our
sample.
}
\label{fig:histoc}
\end{figure}

\section{The sample of EROs}
\label{sec:eros}

\begin{table*}[ht]
\begin{flushleft}
\caption{The Sample of EROs}
\protect\label{tab:eros}
\begin{tabular}{c|c|ccc|ccc|ccc|ccc}
\noalign{\smallskip}
\hline
\noalign{\smallskip}
\multicolumn{1}{c}{} & \multicolumn{1}{c}{area} & \multicolumn{3}{c}{$R-Ks$$\geq$5} & \multicolumn{3}{c}{$R-Ks$$\geq$5.3} 
& \multicolumn{3}{c}{$R-Ks$$\geq$6} & \multicolumn{3}{c}{$R-Ks$$\geq$7} \\
\noalign{\smallskip}
\multicolumn{1}{c}{$Ks$ limits} & \multicolumn{1}{c}{arcmin$^2$}& \multicolumn{1}{c}{N} & \multicolumn{1}{c}{Frac.}  & \multicolumn{1}{c}{Dens.}
& \multicolumn{1}{c}{N} & \multicolumn{1}{c}{Frac.}  & \multicolumn{1}{c}{Dens.}
& \multicolumn{1}{c}{N} & \multicolumn{1}{c}{Frac.}  & \multicolumn{1}{c}{Dens.}
& \multicolumn{1}{c}{N} & \multicolumn{1}{c}{Frac.}  & \multicolumn{1}{c}{Dens.} \\
\noalign{\smallskip}
\hline
\noalign{\smallskip}

$Ks\leq$17.0 &701 & 2 & 0.006 & 0.003& -- & -- & -- & -- & -- & -- & -- & --  & --  \\
$Ks\leq$17.5 &701 & 15 & 0.025 & 0.02& 5 & 0.008 & 0.007& -- &--  &--  &--  & -- & -- \\
$Ks\leq$18.0 &701 & 58 & 0.051 & 0.08& 19 & 0.017 & 0.027& -- &  --& -- & -- & -- & -- \\
$Ks\leq$18.5 &701 & 158 & 0.084 & 0.23& 75 & 0.040 & 0.11& 7 & 0.004 & 0.01& -- &  -- & -- \\
$Ks\leq$18.8 &701 & 279 & 0.111 & 0.40& 150 & 0.060 & 0.21& 30 & 0.012 & 0.04 & 2 & 0.0008 & 0.003\\
$Ks\leq$19.0 &447.5 & 220 & 0.116 & 0.49& 133 & 0.070 & 0.30& 33 & 0.017 & 0.07& 4 & 0.0021 & 0.009\\
$Ks\leq$19.2 &447.5 & 281 & 0.126 & 0.63& 173 & 0.079 & 0.39& 44 & 0.020 & 0.10& 5 & 0.0023 & 0.011\\
\noalign{\smallskip}
$Ks\leq$19.2 & 701+447.5 & 279+119 & 0.127 & 0.67& 150+87 & 0.076 & 0.40& 30+27 & 0.018 & 0.10& 2+4 & 0.0019 & 0.012\\

\noalign{\smallskip}
\hline
\end{tabular}
\end{flushleft}
\footnotesize{We present in detail the cumulative
number (N) of EROs selected at each $Ks$ limiting magnitude, the fraction 
of EROs with respect to the whole field galaxies (Frac.) and the corresponding
surface density (Dens., in objects/arcmin$^2$). {The last line was calculated using the whole survey
area to $Ks=18.8$ and the deeper area to $Ks=19.2$. The data presented here 
are used throughout the paper.}}
\end{table*}

\begin{figure}[ht]
\resizebox{\hsize}{!}{\includegraphics{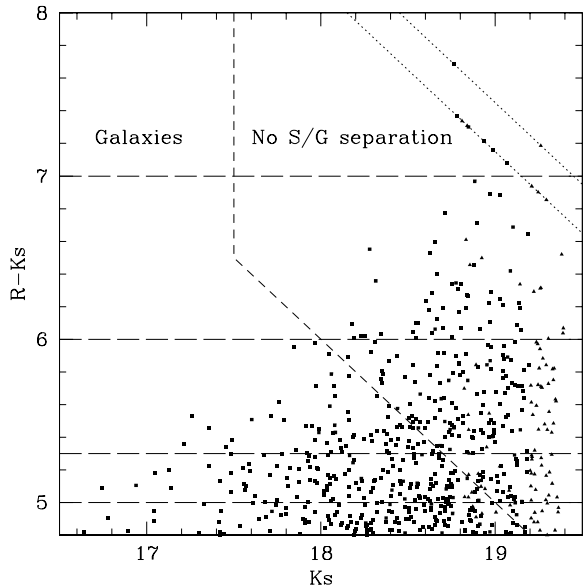}}
\caption{\footnotesize Enlarged portion of Fig. \ref{fig:ccdiag} (bottom) 
around the reddest colors. 
Filled boxes are objects  within the completeness limits of our survey ($Ks\leq19.2$ in the
deeper area and $Ks\leq$18.8 outside) while filled triangles are objects  fainter than those
limits.
The horizontal long-dashed lines correspond to the limits reported
in Table \ref{tab:eros} for the selection of the samples of EROs.
In the region above the short-dashed line the
star--galaxies separation is not feasible. The objects along the diagonal 
dotted lines are not detected in $R$ and have a 3$\sigma$ limit in that band.
}
\label{fig:zoom}
\end{figure}
In Fig. \ref{fig:ccdiag} the $R-Ks$ vs. $Ks$ color - magnitude diagram is
plotted 
for both stars and galaxies in our sample. {
The diagonal straight line in the upper panel indicates
our S/G classification limit in this plane. Because objects
above this line are not classified, no star can appear in the upper right
corner. However, this figure shows that there are very few stars with a 
color redder than $R-Ks > 5$, even in the region
of this color - magnitude diagram where our morphological classification
is still reliable. This suggests that very few stars should be present in the
sample of objects for which no morphological classification was possible.}

Fig. \ref{fig:histoc} shows 
the distribution of colors for galaxies fainter than $Ks=16.5$, from which 
the median color at different $Ks$ magnitude levels have been calculated (see Table 
\ref{tab:median}). 
{
In spite of the presence of objects with color upper and lower limits,
the use of the median colors (instead of the mean colors) allows unbiased 
estimates in the $Ks$ range we are considering.}
The faintest bin with 18.8$<Ks\leq$19.2 
includes only the objects detected in the deeper region. The median $R-Ks$ 
color of the galaxies increases by 0.5 magnitudes from $Ks=16.5$ to $Ks\sim18$ 
and then it remains almost constant up to the limits of our survey ($Ks$=19.2). 
This trend is similar to what is found by Saracco et al. (1999) for the median 
$J-Ks$ galaxies color, which also reaches a maximum at $Ks\sim$ 18--19 and 
then it becomes bluer, while the median $B-K$ color gets significantly
bluer at brighter ($K\sim17$) magnitudes (Gardner et al. 1993).

Our wide-field survey allows us to select a statistically significant
and complete sample of EROs which can provide stringent constraints 
on the density of high-$z$ ellipticals (see Sect. 1).
For this reason, EROs are defined as objects with $R-Ks\geq5$ 
because this corresponds to select passively evolving ellipticals at $z \simgt 0.9$ 
(see Fig.  \ref{fig:color_fig}). Table \ref{tab:eros} shows the results 
of the selection for different magnitude and color thresholds. 
Among the others, the threshold $R-Ks\geq5.3$ is used because it corresponds 
to the selection of $z \simgt 1$ elliptical galaxies (see Fig. \ref{fig:color_fig}). 
We detected 279 EROs with $Ks\leq18.8$ from the whole area and 119 EROs
with $18.8<Ks\leq19.2$ in the deeper area, yielding a total sample of 398 
objects (see Fig. \ref{fig:zoom} and Table \ref{tab:eros}). This is by far the largest sample of EROs obtained to date.  
A small but complete sample of EROs with $R-Ks\geq7$ has also been
selected, and we estimate for the first time their surface density to be
$\sim0.01$ arcmin$^{-2}$ at $Ks\sim19$. 

A comparison of the surface densities of the EROs in our sample with those
in the Thompson et al. (1999) survey can be directly done after taking into 
account the different filters used in the two surveys. 
For the $K^\prime$ filter used 
by Thompson et al. (1999), $Ks$ $\sim$ $K^\prime$ - 0.2 (adopting
$H-K=1$), and therefore their limit at $K^\prime\leq 19$ corresponds 
to $Ks\leq 18.8$ which is the shallower limit of our survey, while their 
$R-K^\prime$ color is bluer than our $R-Ks$ by about 0.1 magnitudes for the
redder objects (Thompson, private
communication). At the level of $Ks\leq18.8$ we find a density of 
$0.042\pm0.008$ arcmin$^{-2}$ for EROs with $R-Ks\geq6$, to be 
compared with the value of 0.039$\pm$0.016 that they find (these errors 
are poissonian). Thus, the two surface densities are in excellent 
agreement with each other.
We also verified that the average $R-Ks$ color of all our objects with
$17.8<Ks<18.8$ ($R-Ks=3.70\pm0.03$, determined with a  Kaplan-Meier
estimator), is in good agreement with the average $R-K^\prime=3.73\pm0.04$ in
$18<K^\prime<19$ by Thompson et al. (2000).

\section{The angular correlation functions}
\label{sec:clustering}

Statistical measurements of the clustering of faint galaxies are 
important for studying the evolution of galaxies and the formation of 
structures in the Universe. In fact, the amplitude of clustering in 
2D space is a useful probe of the underlying 3D structure (e.g. 
Connolly et al. \cite{connolly}, Efstathiou et al. 1991, Magliocchetti 
\& Maddox 1999). The clustering of galaxies on the sky has been studied 
extensively especially in the optical, but also in the near-infrared (e.g. 
Roche et al. 1998 and 1999, Postman et al. 1998, Baugh et al. 1996b). 
Our survey, as noted before, is the widest at the limits of $Ks\sim19$.
It is therefore interesting to estimate the clustering of our sample of
galaxies.

\subsection{Calculation technique}
\label{calculus}

The angular two-point correlation function $w(\theta)$ is defined as 
the excess probability (over a poissonian distribution) of finding 
galaxies separated by the apparent distance $\theta$:
\begin{equation}
dP = N^2[1+w(\theta)]d\Omega_{\rm 1}d\Omega_{\rm 2}
\label{eq:dP}
\end{equation}
where N is the mean density per steradian (Groth \& Peebles, 1977). 

Several methods for estimating $w(\theta)$ from a set of object 
positions have been proposed and used, but the most bias--free and 
suitable for faint galaxies samples resulted to be the Landy \& Szalay 
technique (Landy \& Szalay 1993, see also Kerscher et al. 2000). 
This technique (adopted for the calculations in this paper) 
consists in deriving the counts of objects binned in logarithmic distance 
intervals, for the data-data sample $[DD]$, the data-random sample $[DR]$ 
and the random--random sample $[RR]$. These counts have to be normalized, 
i.e. divided for the total number of couples in each of the 3 samples. 
From them we can estimate $w_{\rm b}(\theta)$ as:
\begin{equation}
w_{\rm b}(\theta) = \frac{[DD]-2[DR]+[RR]}{[RR]}
\label{eq:w_b}
\end{equation}
which is biased to lower values with respect to the real correlation 
function $w(\theta)$:
\begin{equation}
w(\theta) = w_{\rm b}(\theta) + \sigma^2
\end{equation}
where $\sigma^2$ is the "integral constraint" (Groth 
\& Peebles, 1977):
\begin{equation}
\sigma^2 = \frac{1}{\Omega^2} \int\int w(\theta) d\Omega_{\rm 1}d\Omega_{\rm
2}
\end{equation}
Assuming that the angular correlation function $w(\theta)$ can be 
described by a power law of the form $w(\theta) = A \theta^{-\delta}$, then, 
following Roche et al. (1999), we can extimate the ratio between $\sigma^2$  
and the amplitude $A$ using the random--random sample:
\begin{equation}
C = \frac{\sigma^2}{A} = \frac{\sum N_{\rm rr}(\theta)\theta^{-\delta}}{\sum N_{\rm rr}(\theta)} 
\label{eq:C}
\end{equation}

The amplitude of the real two-point correlation function $w(\theta)$ can 
then be estimated by fitting to the measured $w_{\rm b}(\theta)$ the function:
\begin{equation}
w_{\rm b}(\theta) = A(\theta^{-\delta}-C)
\label{eq:correl}
\end{equation}

The errors can be estimated, following Baugh et al. (1996b), as:
\begin{equation}
\delta w_{\rm b}(\theta) = 2\ \sqrt[]{(1+w_{\rm b}(\theta))/DD)}
\label{eq:error}
\end{equation}
where $DD$ is the non normalized histogram of $[DD]$. Eq. (\ref{eq:error}) is 
equivalent to assuming 2$\sigma$ poissonian errors for the correlations, and 
it gives estimates that are comparable to the errors obtained with the 
bootstrap technique (Ling, Frenk \& Barrow 1986). This is necessary because 
it is known that, as the counts in the different bins are not completely 
independent, assuming the 1$\sigma$ poissonian errors would result in
an underestimate of the true variance of the global parameters of the 
angular correlation (see Mo et al. 1992). 

In case of the presence of a randomly distributed spurious component among 
the analyzed sample of objects (an example of this case would be a 
residual stellar component among the galaxy sample), the resulting 
amplitudes are apparently reduced by a factor $(1-f)^2$, where $f$ is 
the fraction of the randomly distributed component (see e.g. Roche et al. 
1999), and the corresponding correction should be applied.

\begin{figure}[h]
\resizebox{\hsize}{!}{\includegraphics{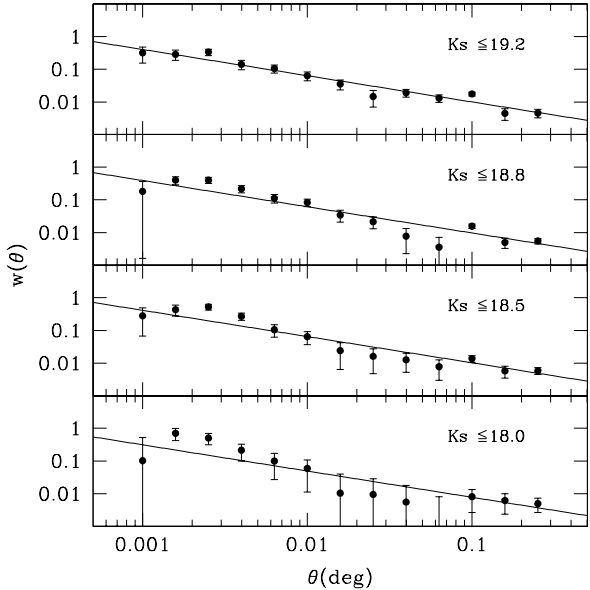}}
\caption{\footnotesize The observed,  bias--corrected two-point 
correlation functions of our $K$--selected sample. The lines plotted 
are the best--fitted power laws (see Table \ref{tab:clust_all}). 
The error in each bin is two times the poissonian error (see Sect. 
\ref{calculus}).
}
\label{fig:all_clust}
\end{figure}

\begin{table}[ht]
\begin{flushleft}
\caption{Clustering amplitudes for the $K$--selected sample}
\protect\label{tab:clust_all}
\begin{tabular}{cccccc}
\noalign{\smallskip}
\hline
\noalign{\smallskip}
\multicolumn{1}{c}{$Ks$ limit}  & \multicolumn{1}{c}{area} & 
\multicolumn{1}{c}{Galaxies} & \multicolumn{1}{c}{A[10$^{-3}$]} & 
\multicolumn{1}{c}{$\delta$} & \multicolumn{1}{c}{C} \\
\noalign{\smallskip}
\hline
\noalign{\smallskip}
18.0 & 701   & 1131 & 1.3$\pm$0.5 &0.8 & 4.55 \\
18.5 & 701   & 1890 & 1.6$\pm$0.3 &0.8 & 4.55 \\
18.8 & 701   & 2503 & 1.5$\pm$0.2 &0.8 & 4.55 \\
19.2 & 447.5 & 2222 & 1.6$\pm$0.2 &0.8 & 5.16 \\
\noalign{\smallskip}
\hline
\smallskip
\end{tabular}
\end{flushleft}
\end{table}

The random samples used in our analysis were obtained using the 
pseudo--random number generator routine of the C Library function 
{\it drand48}. Random samples with up to 200\,000 objects were used. 
Typically the number of objects in the random samples were a factor of 
100--200 larger than the number of observed objects. The random sample 
was generated with the same geometrical constraints as the data 
sample, avoiding for instance to place objects in the regions excluded 
around the brightest stars.

\subsection{The clustering of the $K$-selected field galaxies}

\begin{figure}[ht]
\resizebox{\hsize}{!}{\includegraphics{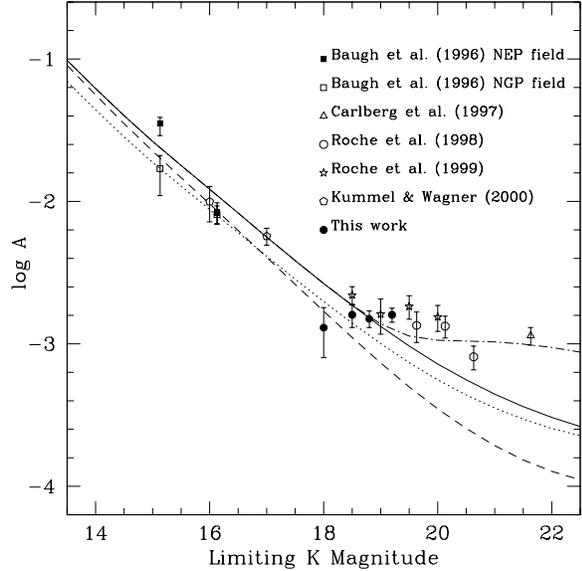}}
\caption{\footnotesize 
The clustering amplitudes measured in our survey, compared with published data from
the literature. The models shown here are from Roche et al. (1998) and Fig. 7 of Roche et al. (1999).
}
\label{fig:roche}
\end{figure}

In our analysis a fixed slope of $\delta=$\ 0.8 was assumed, as this is 
consistent with the typical slopes measured in both faint and bright 
surveys (e.g. Baugh et al. 1996b, Roche et al. 1996, Maddox et al. 1990), and 
because it gives us the possibility to directly compare our results with the 
published ones that are typically obtained adopting such a slope.
The factor $C$ was estimated (with Eq. (\ref{eq:C})) for both the 
whole and the deeper areas, turning out to be 4.55 and 5.16 respectively 
(the angles are expressed in degrees, if not differently stated). 
In Fig. \ref{fig:all_clust} the observed, bias corrected, two-point 
correlation functions $w(\theta)$ are shown; the bins have a constant
logarithmic width ($\Delta\log\theta=0.2$), with the bin centers ranging from 
3.6$\arcsec$ to 15$\arcmin$. 

We clearly detect a positive correlation signal for our sample with an 
angular dependence broadly consistent with the adopted slope $\delta=$\ 0.8,
even if the measurements show some deviations, in particular for the brightest samples.
A few cluster candidates are present in our survey. These possible
clusters include galaxies with $R-Ks\leq 4.5$, and are therefore expected to 
be at $z\simlt0.6$. A detailed analysis of the cluster candidates
will be given in a forthcoming paper. For the purpose of the present
work, we tested that the measured clustering amplitudes are stable in 
case of removal of the galaxies of the most evident cluster from the sample. 
However, the presence of such clusters, most of which happen to be in the 
shallower area, could partly explain the 
observed deviations from the fitted $w(\theta)=A\ \theta^{-0.8}$ power laws
for the three brightest samples. 

\begin{figure}[h]
\resizebox{\hsize}{!}{\includegraphics{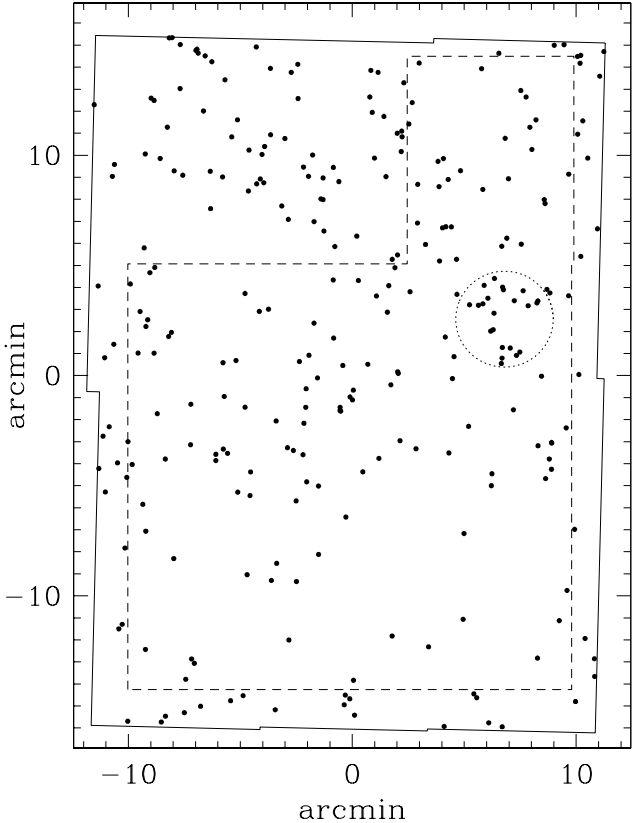}}
\caption{\footnotesize The sky positions of the objects with $Ks\leq$18.8 and $R-Ks$$\geq$5
in our survey. The region enclosed within the dashed line is the deeper region. The colors of the
objects inside the circle are shown in Fig. \ref{fig:eros_cm}.
}
\label{fig:distrib_sky}
\end{figure}

The derived clustering amplitudes are presented in Table \ref{tab:clust_all}. 
The amplitude errors are obtained from the fit assuming Eq. (\ref{eq:error}).
No correction for the stellar contamination was applied.
In Fig. \ref{fig:roche} the clustering amplitudes of 
our samples are compared with other published measurements. The data are shown 
together with a number of PLE models with different clustering evolutions,
which are described in detail in Roche et al. (1998, 1999). 
Our measurements are in good agreement with both the models and the previous 
estimates of Roche et al. (1999), except for the point with limiting
magnitude $Ks=18.0$, which is however the most uncertain of our data points.

As a check it was verified that the correlations of the stellar sample are
consistent with zero, within the measured errors, at all the scales. 
This is a confirmation that the stars are homogeneously distributed on the 
field (as they should be) and the seeing variations across the area did not 
cause a detectable bias in our classification.

\subsection{The clustering of the extremely red objects}
\label{sec:clust_eros}

\begin{table*}[t]
\begin{flushleft}
\caption{Clustering amplitudes for the Extremely Red Objects}
\protect\label{tab:clust_eros}
\begin{tabular}{cccccccc}
\noalign{\smallskip}
\hline
\noalign{\smallskip}
\multicolumn{1}{c}{}  & \multicolumn{1}{c}{area} & 
\multicolumn{2}{c}{$R-Ks\geq5$ sample} & 
\multicolumn{2}{c}{$R-Ks\geq5.3$ sample} & \multicolumn{1}{c}{} & 
\multicolumn{1}{c}{} \\
\multicolumn{1}{c}{$Ks$ limit}  & \multicolumn{1}{c}{arcmin$^{2}$} & 
\multicolumn{1}{c}{Galaxies} & \multicolumn{1}{c}{A[10$^{-3}$]} & 
\multicolumn{1}{c}{Galaxies} & \multicolumn{1}{c}{A[10$^{-3}$]} & 
\multicolumn{1}{c}{$\delta$} & \multicolumn{1}{c}{C} \\
\noalign{\smallskip}
\hline
\noalign{\smallskip}
18.0 & 701   & 58 & 24$\pm$10 & -- & -- &0.8 & 4.55 \\
18.25 & 701   & 106 & 25$\pm$5 & -- & -- &0.8 & 4.55 \\
18.5 & 701   & 158 & 22$\pm$3 & -- & --  &0.8 & 4.55 \\
18.8 & 701   & 279 & 14$\pm$2 & 150 & 14$\pm$3.4 &0.8 & 4.55 \\
19.2 & 447.5 & 281 & 13$\pm$1.5 & 173 & 12$\pm$2.3 &0.8 & 5.16 \\

\noalign{\smallskip}
\hline
\smallskip
\end{tabular}
\end{flushleft}
\end{table*}

The large sample of EROs derived from our survey allowed us for the
first time to estimate their clustering properties. Even a simple
visual inspection of the sky distribution of the objects with $R-Ks\geq5$
(see Fig. \ref{fig:distrib_sky}) shows that the EROs have a very inhomogeneous distribution.

\begin{figure}[ht]
\resizebox{\hsize}{!}{\includegraphics{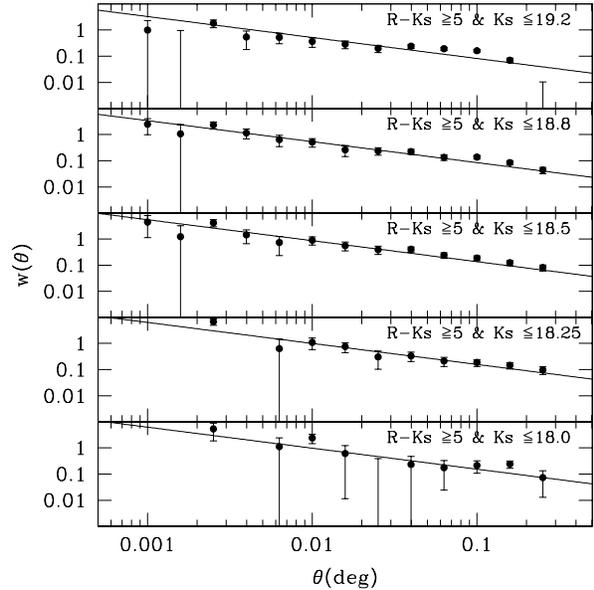}}
\caption{\footnotesize 
The observed, bias--corrected two-point correlations for the sample of EROs (with
$R-Ks$$\geq$5) in our survey.
As in Fig. \ref{fig:all_clust}, the error in each bin is twice 
the poissonian one.
Because of the small number of objects included, some bins for the two
brightest samples, at small angular separation, were not populated. 
The corresponding upper limits would be in agreement with the fitted
amplitudes.
}
\label{fig:eros_clust}
\end{figure}

The results of the quantitative analysis of the clustering are shown
in Fig. \ref{fig:eros_clust}, where the observed, bias--corrected angular
correlations $w(\theta)$ of the objects with $R-Ks$$\geq$5 are displayed.
A strong clustering is indeed present at all the scales that could be
measured, and its amplitudes (Table \ref{tab:clust_eros}) are about an order of magnitude higher than
the ones of the field population at the same $Ks$ limits. 
The correlations are very well fitted by a $\delta=0.8$ power law.
No attempt was made to correct the amplitudes for the stellar 
contamination (see Sect. \ref{calculus}), and we stress that such 
corrections would increase them. Adopting the errors derived from the fits, 
our detections are significant at more than 7$\sigma$ level for the samples 
with $R-Ks\geq5$ and $Ks$ limits equal to or fainter than $Ks = 18.5$. 

The amplitudes shown in Fig. 5 suggest a possible
trend of decreasing strength of the clustering for 
fainter EROs: the $Ks\leq$19.2 EROs are 
less clustered than the ones with $Ks\leq$18.5 and the difference is
significant at 
$2.7\sigma$, based on the derived errors. The significance of this effect
would, however, decrease if the contamination by randomly distributed
stars increases towards the limit of our survey. Although difficult to be
quantified, this is likely to happen
because, differently from the brightest EROs, only a small fraction of the 
EROs fainter than $Ks$ = 18.5 could be morphologically classified
(see Fig. \ref{fig:zoom}).

Defining redder thresholds drastically reduces the number of EROs 
and it is not possible to estimate with sufficient accuracy how 
the amplitudes change for objects with even redder $R-Ks$ colors. We could only
verify that the sample of EROs with $R-Ks\geq5.3$ has clustering 
amplitudes consistent with those of the $R-Ks\geq5$ samples (see Table \ref{tab:clust_eros}).
To measure the clustering of the $R-Ks\geq$6 EROs,
an area at least 10 times larger than ours (i.e. $\sim$2 square degrees) at 
$K=19$ would be needed, assuming that their clustering amplitudes are 
similar to those of the EROs with $R-Ks\geq5$.

\begin{figure}[ht]
\resizebox{\hsize}{!}{\includegraphics{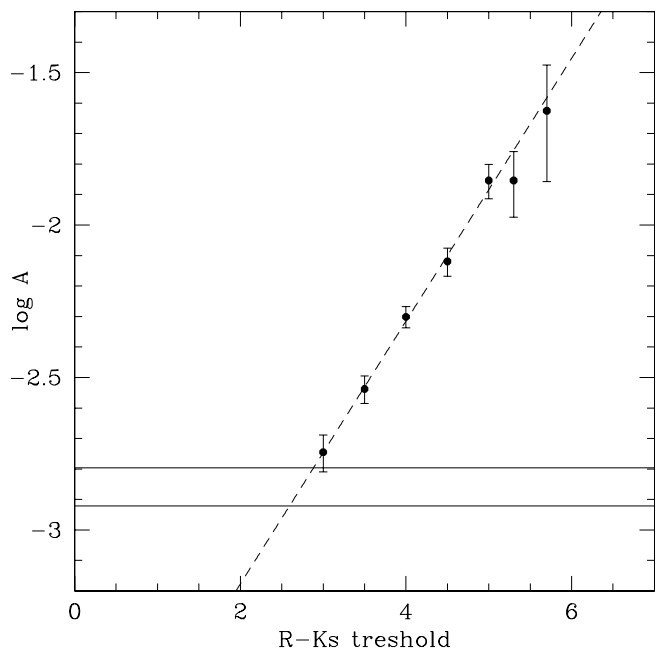}}
\caption{\footnotesize The measured correlation amplitude for the samples
with $R-Ks$ redder than a fixed threshold, in function of the threshold itself. This was
obtained with the $Ks\leq$18.8 sample. The horizontal lines show the $\pm1\sigma$ range for
the clustering of the whole population of field galaxies at this level. 
The fitted power law (dashed line) is $log A = 0.43(R-Ks) - 4.04$.
}
\label{fig:col_var}
\end{figure}

Finally, it was studied if and how the clustering amplitude changes as a 
function of $R-Ks$ for the $Ks\leq$18.8 sample (see Fig. \ref{fig:col_var}). 
A clear increase of $A$ with $R-Ks$ is present for colors $R-Ks$$\geq$3.5, 
while the $R-Ks\geq$3 sample has an amplitude that is consistent with 
that of the whole sample of field galaxies. The variation of $A$ can
be described with a power law in the range of
$3\leq R-Ks \leq5.7$. 
Previous efforts to disentangle the clustering properties of 
the red and blue populations in faint $K$--selected samples probably 
failed because the ERO population was not sufficiently sampled. For 
instance, Kummel \& Wagner (2000) did not find significant differences in the 
clustering of objects with color bluer or redder than $R-Ks$=3.49 
for their $K<$17 sample. This is not surprising since at 
K$<$17 the ERO population is almost absent (see Table \ref{tab:eros} 
and Fig. \ref{fig:ccdiag}).

To check for the stability of these results, possible
systematics that could produce a bias in our work were analyzed. 
{First af all, as the clustering of our $Ks$--selected galaxies is in
good agreement with the literature data (Fig. \ref{fig:roche}), we can
exclude the presence of measurable biases coming from the 
selection of the sample.} 

Regarding the color measuraments, 
since EROs are the tail of objects in the $R-Ks$ color distribution,
systematic variations of the photometric zeropoints across the
area could have the effect 
of creating artificial ERO overdensities and voids. 
To exclude this possibility we verified that the blue tail of the
$R-Ks$ distribution is homogeneously spread across our survey, with a very
low clustering amplitude. In case of zeropoints variations  these should
produce the same effect in both the tails of the color distribution.
Moreover, to test the reality of the large void of EROs clearly seen in the
bottom region of our survey (see Fig. \ref{fig:distrib_sky}),
the $R-Ks$ color distribution  of the galaxies inside and outside
this large void were compared by means of a Kolmogorov-Smirnov test, selecting only the galaxies with
$R-Ks\leq$4 in both regions. The probability that the two distributions are
extracted from the same population is 43\%. {Thus, the two regions 
are fully consistent with each other with respect to the color distribution
of the blue population. This, together with the fact that no underdensity
is present in this region when only the bluer galaxies are considered,
shows that the void of EROs should be considered a real feature.  }

All these tests strongly suggest that
the inhomogeneous ERO distribution is a real effect.

\section{Main implications}
\label{sec:implic}

\subsection{On the nature of EROs}

The strong clustering signal that we find to increase with the $R-Ks$
threshold and to reach very high values for the EROs is potentially
capable of giving insight about the nature of these objects. 

The main possible source of objects that may contribute to the
ERO population, as discussed in the introduction, are old passively
evolving $z\simgt1$ ellipticals, dust-reddened starburst galaxies and,
in case of unresolved objects,
low-mass stars or brown dwarfs. Our field being at high galactic
latitude ($b\sim57\degr$), stars are expected to have no clustering, and to
be homogeneously distributed,  and certainly not to give the strong clustering signal detected.
As for the starburst galaxies, it must be noted that in such galaxies the
red colors are mainly driven by the amount of dust extinction and not by the
redshift, as in the case of ellipticals (see Fig. \ref{fig:color_fig}), and
therefore a wide redshift distribution is expected which should dilute their
intrinsic clustering. Moreover, it is known that the IRAS--selected
galaxies (which are typically star--forming galaxies) have very low 
intrinsic clustering (e.g. Fisher et al. 1994). 
We can therefore reasonably conclude that the observed signal 
is due to the clustering of high  redshift ellipticals.
This is also suggested by  studies of the local universe
which have shown that early-type galaxies are much more clustered than late-type 
galaxies (e.g. Guzzo et al. 1997, Willmer et al. 1998). 
In this regard the results plotted in Fig. \ref{fig:col_var} could be 
qualitatively explained by noting that in selecting redder samples the fraction
of early type galaxies increases (the color of local ellipticals is just 
around $R-Ks\simlt3$) and by assuming that these galaxies are intrinsically 
more clustered, while such plot would be difficult to understand if
mainly driven by the strongly reddened starburst galaxies.
These considerations strongly suggest that EROs are mainly
composed by $z\simgt1$ ellipticals, confirming
the previous indications that had been found on this issue.

\begin{figure}[ht]
\resizebox{\hsize}{!}{\includegraphics{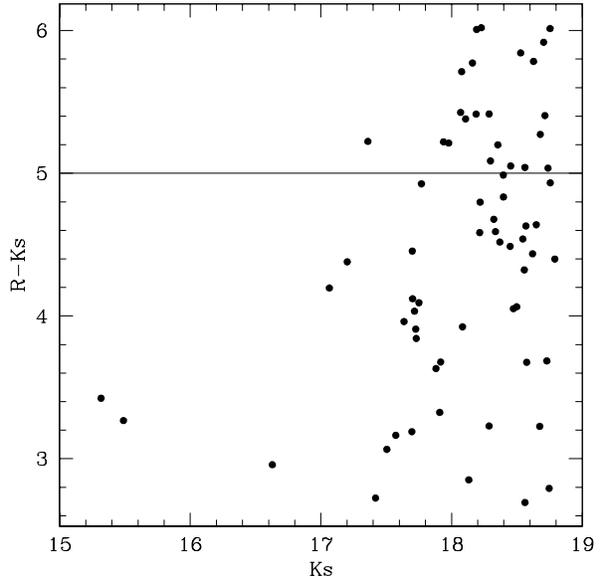}}
\caption{\footnotesize Color magnitude diagram of all the objects with 
$Ks\leq$18.8 in the region inside the circle of Fig. \ref{fig:distrib_sky}.
}
\label{fig:eros_cm}
\end{figure}

As the elliptical galaxies are the dominant population of galaxy clusters, 
we investigated the possibility that the detected clustering of EROs
could be the result of a few massive clusters at $z\geq1$ present in our field. 
For example, in the region inside the circle in Fig. \ref{fig:distrib_sky},
a large ERO overdensity is found, that one could suspect to be due to
a high-$z$ cluster of galaxies. However, Fig. \ref{fig:eros_cm} 
shows that there is no clear color-magnitude sequence among the 
$R-Ks>5$ objects inside that region, suggesting that they do not all
belong to a single cluster. In case of a cluster, even at
high--$z$, a well defined color-magnitude sequence is in fact generally  observed
(e.g. Stanford et al. 1998). 
In the last years a few examples of massive $z\simgt$1 clusters of galaxies have
been discovered (e.g. Stanford et al. 1997, Rosati et al. 1999), with
X ray luminosity of $\sim10^{44}$ erg s$^{-1}$. A crude
estimate of the number of structures of this sort that could be observed
in our survey can be derived by calculating the number of high--$z$ clusters 
with L$_{\rm X}>10^{44}$
erg s$^{-1}$ expected in the volume we are sampling. 
From the X ray luminosity function of such structures at $z\sim1$ 
(Rosati et al. 2000) we estimate that the expected number of massive
clusters in our field in the redshift interval $0.9<z<2$ is only $\sim0.1$ 
(for $\Omega_{\rm 0}=1$). 

Moreover, the detection of the ERO positive correlation, 
following a $\delta=0.8$
power law on all the scales from 10$\arcsec$ to 
15$\arcmin$ (corresponding to $\sim$8 Mpc at $z\sim1$)
suggests that the clustering signal does not come from a 
few possible clusters detected in our field, but rather from the whole 
large scale structure traced by the elliptical galaxies.

\subsection{Fluctuations of the ERO number density}

Our results on the clustering of EROs have important consequences 
on the problem of estimating the density of high-$z$ ellipticals 
(see Sect. 1).

\begin{figure}[ht]
\resizebox{\hsize}{!}{\includegraphics{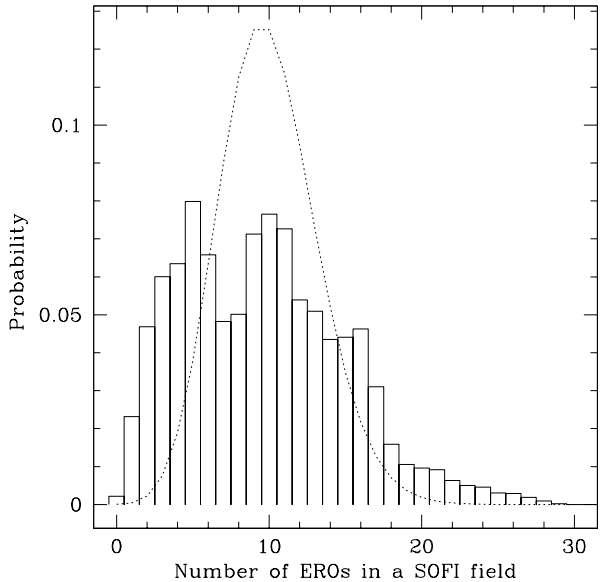}}
\caption{\footnotesize The histogram shows the frequency distributions of the number of EROs ($R-Ks$$\geq$5 and
$Ks\leq18.8$) that have been recovered by sampling our survey with a SOFI field (25 arcmin$^2$). The mean
expected value is 10. The  dotted curve represents the probability distribution
expected in the poissonian case.
}
\label{fig:SOFI_f}
\end{figure}

The existence of an ERO angular correlation with $\delta=0.8$
and with a high amplitude implies significant surface 
density variations around the mean value 
even for relatively large areas.
In the presence of a correlation with amplitude A, the rms fluctuations 
of the counts around the mean value $\overline{n}$ is (see for example Roche et 
al. 1999):
\begin{equation}
\sigma_{true}^2 = \overline{n}\ (1+\overline{n}AC)
\label{eq:sigma}
\end{equation}
The factor $C$ is the same as in Eq. (\ref{eq:C}) and, by applying Eq.
(\ref{eq:C}) for several areas, it was found that it can be approximated
as:
\begin{equation}
C = 58\ {\rm Area}^{-0.4}
\label{eq:C2}
\end{equation}
if the area is expressed in arcmin$^2$ and $\delta=0.8$. 
The validity of such an approximation has been tested for square regions and for areas not
larger than the ones of our survey.
With Eq. (\ref{eq:sigma}) and (\ref{eq:C2}) the expected variations of 
the ERO number counts can be calculated, once their clustering amplitude 
is known.

To verify the consistency of this picture, we derived the distribution 
of the number of EROs (with $R-Ks$$\geq$5 and $Ks\leq$18.8, i.e. those in Fig.
\ref{fig:distrib_sky}) that can be recovered in our area by sampling it with a
field of view of 5$\arcmin\times$5$\arcmin$, which is the typical field of 
view of a near-infrared imager such as SOFI. In Fig. \ref{fig:SOFI_f} 
the observed frequencies of the number of EROs recovered in this 
counts-in-cell analysis is plotted. As the mean expected number of EROs 
is about 10, the 
poissonian fluctuations would be $\sigma_{\rm poisson}\sim$3.2, while 
fluctuations with $\sigma=$5.4 are actually observed. 
Applying Eq. (\ref{eq:sigma}), 
the measured clustering amplitude $A=0.013$ implies $\sigma=5.55$, in excellent
agreement with the measured $\sigma$ value.

We also note that the distribution of the numbers of EROs in Fig. 
\ref{fig:SOFI_f} is not only asymmetric, but also very broad, ranging from $N$=0 to 
$N$=30. In 29\% of the cases the number of EROs recovered is $N\leq$5, 
corresponding to a surface density half of the real one, while only in 19\% of 
the cases the observed number is $N\geq$15. This shows
that, 
on average, it is more 
probable to underestimate the real surface density of these objects.
This is a clear property of the sky distribution that we observe, as
the voids extend on a large fraction of the surveyed area.
These results show how strong the effects of the field-to-field
variations are in the estimate of the sky surface
density of EROs. In this respect, it should be noted here that all previous 
estimates of the number density of high--$z$ ellipticals
were based on surveys 
made with small fields of view, typically ranging from 1 arcmin$^2$ 
in the case of the NICMOS HDF-S (Benitez et al. 1999) to 60 arcmin$^2$ 
in the case of Barger et al. (2000).

\subsection{Implications for the evolution of elliptical galaxies}

\begin{figure}[ht]
\resizebox{\hsize}{!}{\includegraphics{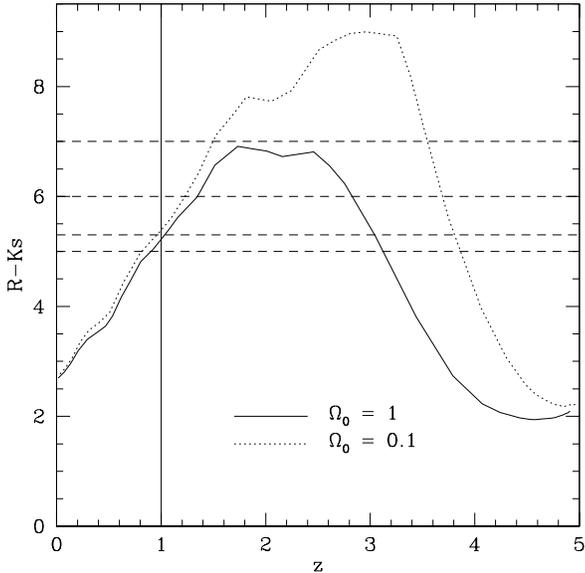}}
\caption{\footnotesize The color--redshift relation for the extreme PLE model described in
the test, computed for a flat and an open universe ($\Lambda=0$). 
The horizontal dashed lines are the thresholds adopted in this paper. 
The vertical line shows that $z>1$ ellipticals correspond to
the $R-Ks>5.3$ EROs. At $z=1$ a $Ks=19$ elliptical galaxy has $L\sim0.3 L_{\rm
*}$ and
$L\sim0.5 L_{\rm *}$ in our flat and open models, respectively.
}
\label{fig:color_fig}
\end{figure}

The selection of galaxies with colors $R-Ks>5$ can be used to search 
for elliptical galaxies at $z>0.9$ (see Fig. \ref{fig:color_fig}), and 
to study their evolution by comparing their observed surface densities 
with those expected from PLE or hierarchical models of massive galaxy 
evolution. In this respect, very discrepant results have been obtained 
so far, making the formation of spheroids one of the most controversial 
problems of galaxy evolution (see the Introduction). 

Our results on the ERO clustering clearly show that for such a comparison 
to be reliable, both a wide field survey (resulting in a large number of 
EROs) and a consistent estimate of their surface density fluctuations 
are necessary before reaching solid conclusions on the evolution
of elliptical galaxies. 

In this section, 
{with the main aim to show the effect of the increased uncertanties due to the clustering,}
a preliminary comparison is presented between the
sky density of EROs observed in our survey (Table \ref{tab:eros}) and the predictions of an 
extreme PLE model similar to that used by Zepf (1997).
In this model, ellipticals formed at $z_{\rm f}$=5 and their star formation 
rate ($SFR$) is characterized by an exponentially decaying burst with 
$SFR\propto exp(t/\tau)$, with $\tau=0.1$ Gyr. Adopting the Markze et 
al. (1994) local luminosity function of ellipticals, and the Bruzual \& 
Charlot (1997) models with solar metallicity and Salpeter IMF,  
the expected surface densities of passively evolving ellipticals with $R-Ks\geq6$
was calculated  for different limiting $Ks$ magnitudes.

\begin{figure}[ht]
\resizebox{\hsize}{!}{\includegraphics{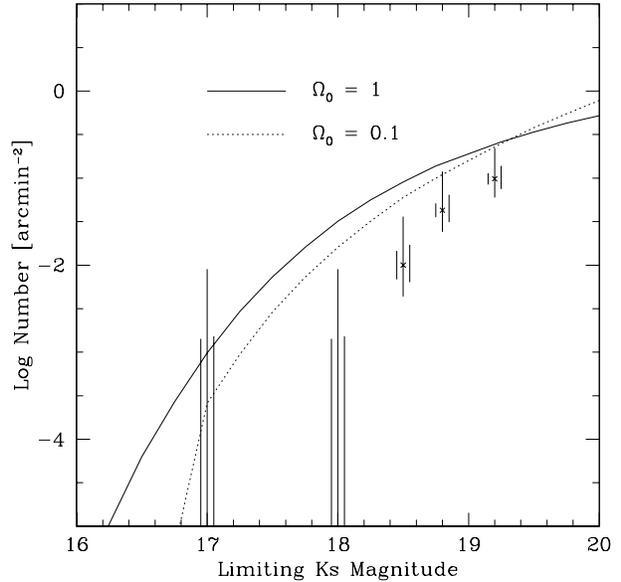}}
\caption{\footnotesize The observed integral number densities of EROs with $R-Ks$$\geq$6 
(Table \ref{tab:eros}) are compared with the extreme PLE model described in the text. The three error 
bars shown here are, from left to right, the 1$\sigma$ poisson uncertainty and 
the 2$\sigma$ and 1$\sigma$ {\it true} uncertanties. The {\it true} uncertanties 
take into account the observed clustering of EROs. 
}
\label{fig:eros_dens}
\end{figure}

Fig. \ref{fig:eros_dens} shows the comparison between the expected and the
observed densities of EROs with $R-Ks\geq6$ (such a color
threshold should select passively evolving galaxies at
$z\simgt1.3$).
For each data point we show three different error bars, which are actually the
region of confidence in the poissonian case (at 1$\sigma$) and in the true
(i.e. clustering corrected) case (at 1$\sigma$ and 2$\sigma$).
Such confidence regions have been estimated, following the 
prescriptions of Eq. (\ref{eq:sigma}), by finding the range of values 
for the true average counts $\overline{n}$ for which the observed $N$ would 
represent a deviation of the required number of $\sigma$ from the
real density. In other words such ranges are defined from the two 
solutions of the equation:
\begin{equation}
\alpha^2 = \frac{(\overline{n}-N)^2}{\overline{n}(1+\overline{n}AC)}
\label{eq:chi2}
\end{equation}
where $\alpha$ is the number of $\sigma$ considered.
The amplitudes of the angular correlation function
used for the $R-Ks\geq6$ EROs are those derived for the 
$R-Ks\geq5$ EROs, which is likely to be a conservative assumption as the amplitudes of 
redder samples should be higher, as suggested by Fig. \ref{fig:col_var}.

Fig. \ref{fig:eros_dens} shows that the observed EROs densities are indeed
lower than the predictions of this particular PLE model. However, even 
for the most deviant point, 
this PLE model can be rejected at only the 2.5$\sigma$ and 2.3$\sigma$
level for $\Omega_{\rm 0}=1$ and $\Omega_{\rm 0}=0.1$ respectively, if we use the
``correct'' error bars. Note that the 
different points plotted in that figure are not statistically
independent because they are partially obtained with the same 
objects (they are cumulative values).

It should be recalled here that the observed ERO densities plotted
in Fig. 14 are an overestimate of the true density of the ellipticals 
because of the contamination by dusty starbursts (see Cimatti et al.
1998, 1999; Dey et al. 1999; Smail et al. 1999) and by field
low-mass stars. The fraction of dusty starbursts in complete ERO samples 
is not known yet, as discussed in the introduction, but our results show
that they should not be the dominant population.
For instance, assuming that
the fraction of dusty starbursts and low-mass stars is 20\% and 10\% 
of the ERO population respectively, this would decrease the observed 
densities plotted in Fig. \ref{fig:eros_dens} accordingly, but it
would also increase by a factor of 2 the clustering amplitudes of 
the high-$z$ ellipticals (see Sect. \ref{calculus}), and hence the error 
bars related to those points.  As a consequence, the statistical
significance of the difference between data and model in this case
would be only at the 2$\sigma$ level.

It is relevant to mention that the predictions of the PLE models
depend very strongly on many parameters that have to be adopted 
{\it a priori} such as $H_{\rm 0}$, $\Omega_{\rm 0}$, the local LF of ellipticals
(uncertain by up to a factor of 2), the redshift of formation $z_{\rm f}$, 
the history of star formation, the metallicity, the IMF, the 
spectral synthesis models. For instance, even just a decrease of 
$z_{\rm f}$, or a small residual star formation at z $\sim$ 1.5
(Menanteau et al. 98,
Jimenez et al. 99), would decrease the predicted numbers of EROs making them 
more consistent with 
our data. We therefore conclude that it seems premature to reject even
extreme PLE models at a high level of statistical significance on the basis
of these data.

A preliminary comparison of our results can be made with 
some aspects of the hierarchical models of galaxy formation. First of
all, our findings could qualitatively fit into the predictions of such models, 
where high-$z$ ellipticals should be very clustered (Kauffmann et al. 1999) 
because they are expected to be linked to the most massive dark matter 
haloes which are strongly clustered at high--$z$. The indication (marginally
significant at $\sim2.7\sigma$ level) of a decrease of the clustering amplitude 
of the EROs with the $Ks$ magnitude (see Sect. \ref{sec:clust_eros}), 
if mainly due to the mass of the galaxies, 
could also fit well in this framework because smaller objects should be 
connected to smaller dark matter haloes which are expected to be less 
correlated. 
On the other hand, our results seem to conflict
with the predictions made by Kauffmann \& Charlot (1998) on the fraction
of $K$-selected galaxies with $K\leq19$. In fact, the fraction of galaxies 
observed to have color $R-Ks\geq5.3$ (which corresponds to the 
selection of $z \simgt 1$ ellipticals) is about 7\% of the total in our survey 
(see Table \ref{tab:eros}), to be compared with the 2--3\%
of $z>1$ galaxies with K$\leq$19 expected in the Kauffmann \& Charlot 
(1998) hierarchical model.
This result on the fraction of $z\simgt1$ galaxies in our $K$-selected
sample broadly agree with the finding of Eisenhardt et al. (2000).

\section{Summary}

The main results of this work are:

\begin{itemize}

\item[--] We have presented a survey which covers 701 arcmin$^2$  
and is 85\% complete to $Ks\leq18.8$
over the whole area and to $Ks\leq19.2$ over 447.5 arcmin$^2$; the R-band
limit is $R\geq26.2$ at the 3$\sigma$ level. 

\item[--] The observed galaxy counts are derived over the largest area 
so far published in the range of $18\leq Ks\leq19.2$. 
Such counts are  
in excellent agreement with other published data.

\item[--] The median $R-Ks$ color of field galaxies increases by 0.5 mags 
from $Ks=16.5$ to $Ks=18$, and it remains constant to $Ks=19.2$. 

\item[--] A sample of 398 EROs has been selected. This sample is the largest 
published to date and is characterized by an area larger by
about four times than previous surveys. 
The ERO counts and 
surface densities have been derived for several color thresholds and 
$Ks$ limiting magnitudes. In particular, we find 
$0.67\pm0.03$ (poissonian) EROs arcmin$^{-2}$ with  $R-Ks\geq5$ and
$0.10\pm0.01$ EROs arcmin$^{-2}$ with  $R-Ks\geq6$ at  $Ks\leq19.2$. 

\item[--] The surface density of EROs with $R-Ks\geq7$ has been
estimated for the first time to be of the order of $\sim0.01$ arcmin$^{-2}$ at
$Ks\sim19$.

\item[--] The angular correlation function of field galaxies, fitted with a
fixed slope $\delta = 0.8$, has an amplitude $A(1\degr)\sim0.0015$ at 
$18.5\leq Ks\leq 19.2$, in agreement with previous measurements.

\item[--] For the first time, we detected the clustering of EROs, with an
amplitude $A(1\degr)\sim0.015$ for the objects with 
$R-Ks\geq5$, in the range $18.5\leq Ks\leq 19.2$) which is about a factor of 
ten higher than that of field galaxies. The ERO two point correlations are very well
fitted by a $\delta = 0.8$ power law.

\item[--] The clustering amplitude of the galaxies increases 
with the $R-Ks$ color threshold following 
the relation \ $log A \propto 0.43(R-Ks)$, for $3\leq R-Ks\leq 5.7$
at $Ks\leq 18.8$.

\item[--] The strong clustering of EROs is shown to be a direct evidence
that a large fraction of these objects are indeed high--$z$ ellipticals.  
Our result is therefore the first detection
of the large scale structure traced by the elliptical galaxies at $z\sim1$.

\item[--] The ERO clustering explains the conflicting results obtained 
so far on the density of high-$z$ ellipticals in terms of strong 
field-to-field variations affecting the surveys based on small fields 
of view (e.g. $5\times5$ arcmin).

\item[--] Taking into account the clustering of EROs, even
the predictions of extreme PLE models for the comoving density of 
high--$z$ ellipticals cannot be rejected at much more than 2$\sigma$
significance level.

\end{itemize}

\begin{acknowledgements}

We would like to thank Nathan Roche for providing his models in digital 
form, Gustavo Bruzual and Stephane Charlot 
for their synthetic stellar population models. We also thank Leonardo
Vanzi for his assistance during the NTT observations and
the anonymous referee for useful comments.
LP acknowledges the support of CNAA during the realization of this
project.

\end{acknowledgements}

\end{document}